\documentclass[twocolumn,preprintnumbers,amsmath,amssymb,floatfix]{revtex4}

\usepackage{graphicx}
\usepackage{dcolumn}
\usepackage{amsmath}
\usepackage{amssymb}

\makeatletter
\def\btt#1{\texttt{\@backslashchar#1}}
\DeclareRobustCommand\bblash{\btt{\@backslashchar}}
\makeatother

\begin{document}

\title{Quantum Codes for Simplifying Design and Suppressing Decoherence in
Superconducting Phase-Qubits}
\author{Daniel A. Lidar}
\email[Corresponding author. Email: ]{dlidar@chem.utoronto.ca}
\author{Lian-Ao Wu}
\email[Email: ]{lwu@chem.utoronto.ca}
\affiliation{Chemical Physics Theory Group, University of Toronto, 
80 St. George Street, Toronto, Ontario M5S 3H6, Canada}
\author{Alexandre Blais}
\email[Email: ]{ablais@physique.usherb.ca}
\affiliation{D\'epartement de Physique and Centre de Recherche sur les Propri\'et\'es \'Electroniques de Mat\'eriaux Avanc\'es,
Universit\'e de Sherbrooke, Sherbrooke, Qu\'ebec, J1K 2R1, Canada}

\date{\today }

\begin{abstract}
We introduce simple qubit-encodings and logic gates which eliminate the need for certain
difficult single-qubit
operations in superconducting phase-qubits, while preserving
universality. The simplest encoding uses two physical qubits per
logical qubit. Two 
architectures for its implementation are proposed: one employing $N$
physical qubits out of which $N/2$ are ancillas fixed in the $|1\rangle $
state, the other employing $N/2+1$ physical qubits, one of which is a bus
qubit connected to all others. Details of a minimal set of universal encoded
logic operations are given, together with recoupling schemes, that
require nanosecond pulses. A
generalization to codes with higher ratio of number of logical qubits per physical qubits is presented. Compatible
decoherence and noise
suppression strategies are also discussed.
\end{abstract}

\maketitle

\section{\label{intro}Introduction}

Solid state systems are now attracting much attention as potential
components for quantum computers, in part because of their potential
scalability, and because the parameters of a solid state qubit can be engineered
with considerable freedom. This freedom allows one to add different building blocks
to the quantum computer, each block bringing the design closer to
satisfying the criteria required for quantum computation (QC)
\cite{DiVincenzo:00} and, in particular, enabling universal QC
according to the (non fault-tolerant) ``standard paradigm'' \cite{Barenco:95a}. In this standard
paradigm, all single-qubit operations [generating $SU(2)$] plus an
entangling two-qubit operation (e.g., controlled-NOT, denoted {\sc CNOT}) are necessary to achieve universal quantum
computation. Alternatively, one can use a discrete set of single-qubit
operations (e.g., Hadamard and $\pi/8$) together with {\sc CNOT} in order to
approximate to arbitrary accuracy any quantum circuit
\cite{Boykin:99}, a
method that is compatible with fault-tolerant quantum error correction
\cite{Aharonov:96,Knill:98}. While the different building blocks help in reaching
universality, each one comes with its fabrication difficulties and
adds a potential source of noise and decoherence to the system.

It was realized recently that some of these building blocks can be
replaced by `software' means
\cite{Bacon:99a}.
More specifically, by encoding a logical qubit into a few physical
qubits, the design of the quantum computer 
can be simplified without compromising universality. This approach is
known as encoded universality \cite{Bacon:Sydney}. E.g., it is
possible to remove difficult-to-implement single-qubit operations \cite{Bacon:99a,Kempe:00,DiVincenzo:00a,Bacon:Sydney,WuLidar:01,Levy:01,Benjamin:01,LidarWu:01,WuLidar:01a,Kempe:01,Kempe:01a,WuLidar:02,Vala:02}. This simplification
has the advantages of facilitating fabrication and reducing some sources
of decoherence and noise. Interesting alternatives to encoding that
also aim to reduce design constraints, some by replacing logic gates with
measurements, were recently presented in
\cite{Raussendorf:01,Nielsen:01,Leung:01a,Masanes:02}.

Studies of simplifying encodings have so far been performed
primarily for
exchange-type Hamiltonians in spin-coupled solid state quantum
computer designs
\cite{Bacon:99a,Kempe:00,DiVincenzo:00a,Bacon:Sydney,Levy:01,WuLidar:01,Benjamin:01,LidarWu:01,WuLidar:01a,Kempe:01,Kempe:01a,WuLidar:02,WuByrdLidar:02,WuLidar:01b,ByrdLidar:01a,Vala:02},
and in NMR \cite{Viola:01a,Fortunato:01}.
Here we extend this study to superconducting phase qubits.

Superconducting phase qubits are among the leading solid state qubit
candidates, in part due to recent experimental progress \cite{vdWal:00,Friedman:00}. Several designs of phase qubits have been
suggested in the literature; see Ref.~\cite{Makhlin:01} for a review. In
this work, we first focus on the d-wave grain boundary qubits
\cite{Zagoskin:99,Blais:00} and recall the relevant system
Hamiltonian. A simplification of the design suggested in Refs.~\cite{Zagoskin:99,Blais:00}
will yield a system Hamiltonian which is not versatile enough to be
universal, according to the standard paradigm
\cite{Barenco:95a,Boykin:99}, because it lacks certain single-qubit operations. We will
show how universality can be recovered by using a simple encoding and
recoupling techniques. The encoding will suggest the use of a ``bus qubit'', a
concept that could be useful for other quantum computer designs. In
particular, we explore the possibility of using these concepts with the
other superconducting phase qubit designs. Finally, application of the
dynamical decoupling technique to decoherence reduction will be examined.

\section{System Hamiltonian}

\label{system-hamiltonian}

\begin{figure}
\centering
\includegraphics[height=5cm]{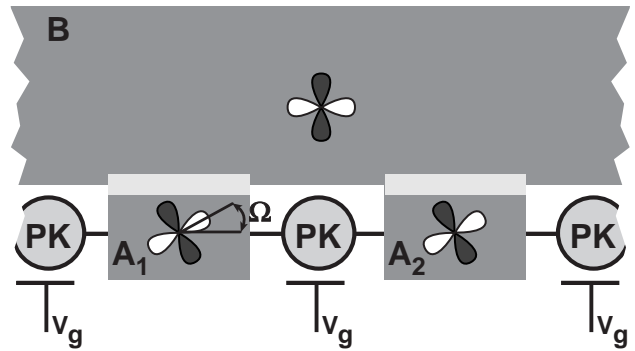}
\caption{Superconducting grain boundary qubits: Terminals A$_1$, A$_2$ and B
are d-wave superconductors, PK (parity key) represent the SSETs, $\Omega$ is
the mismatch angle between the lattices of A$_i$ and B. Positive lobes of
the d-wave order parameter are shaded. Two physical qubits are shown.}
\label{fig_gbqb}
\end{figure}

The d-wave grain boundary qubit (dGB qubit), Figure~\ref{fig_gbqb}, has the
following system Hamiltonian \cite{Blais:00}: 
\begin{equation}
H_{S}=H_{X}+H_{Z}+H_{ZZ},
\end{equation}
where 
\begin{eqnarray}
H_{X} =\sum_{i=1}^{N}\Delta _{i}\sigma _{i}^{x}\quad {\rm tunneling}
\label{H_x}
\end{eqnarray}
\begin{eqnarray}
H_{Z} =\sum_{i=1}^{N}b_{i}\sigma _{i}^{z}\quad {\rm bias}
\label{H_z}
\end{eqnarray}
\begin{eqnarray}
H_{ZZ} =\sum_{i,j=1}^{N}J_{ij}\sigma _{i}^{z}\otimes \sigma _{j}^{z}\quad 
{\rm Josephson\,coupling},
\label{H_zz}
\end{eqnarray}
and where $\sigma_i^\alpha$ ($\alpha = x,y,z$) are the Pauli matrices. In this system, coherent tunneling of the phase is only possible when the
energy levels are in resonance \footnote{Tunneling can actually be possible at finite bias, e.g. due to finite level width. The tunneling amplitude will however be very sensitive
to fluctuations of these biases. Therefore here we only consider
biases large enough to suppress tunneling.}. As a
result, when a bias or
Josephson coupling is turned on, the tunneling matrix element(s) $\Delta _{i}
$ for the corresponding qubit(s) is exponentially suppressed. We therefore take  {\em only one} of the terms (\ref{H_x}), (\ref{H_z}) or (\ref{H_zz}) to be on at any given time.

Moreover, for the dGB qubit, turning on the bias or
Josephson coupling is the only way to control the value of the tunneling
matrix element. The latter can then effectively only be turned on or off,
without continuous control over its magnitude.
This magnitude is determined
at fabrication time by the asymmetry between the d-wave superconductors
forming the qubit and by the width of the junction \cite{Zagoskin:99}. It is
interesting to note that by
connecting an external capacitor to the qubit circuit,
the magnitude of $\Delta _{i}$ could be controlled independently \cite{Blatter:01}. However,
this design complicates fabrication and connects the qubit to another
potential source of decoherence. Since the tunneling matrix elements depend
exponentially on their parameters, small noise on these parameters can have
dramatic effect on the coherence of the system. In this paper, we are
interested in simplifying fabrication demands on the design and reducing
sources of decoherence, hence this possibility will not be considered
further.

We now come to our main simplification: {\em In this work we reduce the constraints on fabrication by removing the
possibility of applying bias $b_{i}$ on individual qubits}. This bias
requires, e.g., the possibility of applying a local magnetic field on
each qubit, and is experimentally very challenging to realize. We do retain the Josephson couplings $J_{ij}$ between
the qubits, as these are necessary to produce entanglement. These couplings
are realized by connecting pairs of qubits by a superconducting single
electron transistor (SSET) \cite{Zagoskin:99,Blais:00}. The magnitude of $
J_{ij}$ can, to some extent, be tuned continuously by the SSET's gate
voltage. The sign of this energy could be changed by inserting a strong
$\pi$-junction.  Again, with simplification of design in mind, the sign of $J_{ij}$
will hereafter be considered fixed.

The effective system Hamiltonian that we consider in this paper is
therefore:
\begin{equation}
H_{S} = H_X + H_{ZZ},  \label{gbqb_hamiltonian}
\end{equation}
where we have {\em continuous} control over $J_{ij}$. As mentioned
above, when the Josephson coupling $J_{ij}$ is non-zero, the corresponding matrix
elements necessarily vanish: $\Delta_i=\Delta_j=0$. In the idle state, $
\Delta_i$ is non-zero and the qubit undergoes coherent tunneling. We
proceed to show how universal QC can be performed given the outlined constraints.

\section{Encoding, Initialization, and Encoded Logic Gates}

\subsection{Code}

In Ref.~\cite{Blais:00}, it was shown how to perform universal quantum
computation on the dGB qubit in the case where individual
control of the bias $b_i$ is possible. Here, we do not allow for this possibility and,
since $H_{X}$ and $H_{ZZ}$ are insufficient for universal QC
on the physical qubits, the techniques of Ref.~\cite{Blais:00} will
have to be supplemented. This is achieved by encoding a pair of physical
qubits as one logical qubit in the following way:

\begin{eqnarray}
|0_{L}\rangle _{m} &=&|0\rangle _{2m-1}\otimes \left|
1\right\rangle_{2m}\equiv |0_{2m-1}1_{2m}\rangle  \nonumber \\
|1_{L}\rangle _{m} &=&|1\rangle _{2m-1}\otimes \left|
1\right\rangle_{2m}\equiv |1_{2m-1}1_{2m}\rangle  \label{eq:encoding}
\end{eqnarray}
for the $m^{{\rm th}}$ logical qubit with $m=1,2,.....,N/2$ and where there
are a total of $N$ physical qubits. It is easy to check that any other
encoding into two qubits is not preserved under
$H_{X}+H_{ZZ}$. Moreover, if $\sigma_{2m}^x$ terms are allowed to act
then the encoding of Eq.~(\ref{eq:encoding}) is not preserved
either. We address this issue below.

\subsection{Initialization}

Initialization of each encoded qubit can be done by measurement. This will
project on the physical qubit's computational basis. Hence measuring the $
(2m-1)^{{\rm th}}$ physical qubit will yield either $|0\rangle _{2m-1}$ or $
|1\rangle _{2m-1}$ (either one would do), while measuring the $(2m)^{{\rm th}}$
physical qubit will have to be repeated until the outcome $
\left|1\right\rangle _{2m}$ is obtained. Since tunneling will cause
bit-flips immediately following the measurement, a good strategy is to
simultaneously measure both these physical qubits, and repeat the
measurement of the $(2m-1)^{{\rm th}}$ physical qubit (projecting on the
computational basis every time) until that of the $(2m)^{{\rm th}}$ physical
qubit has converged on $\left| 1\right\rangle _{2m}$. Alternatively, initialization
can be performed by field-cooling the qubits.  By choosing the proper orientation
for the external field, all physical qubits can be prepared  in the $|1\rangle$
state and, therefore, all logical qubits in the $|1_L\rangle$ state.  This strategy
will most probably not be of practical use during computation but can serve
to provide an initial supply of fresh qubits.  For the dGB qubits, a current in terminal
B can also be used to initialize all qubits to the $|1_L\rangle$ state \cite{Zagoskin:99}.

\begin{figure}
\centering
\includegraphics[height=12cm,angle=270]{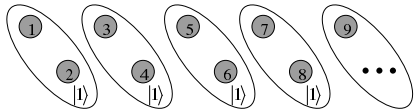}
\vspace{-7cm}
\caption{Triangular arrangement of physical qubits, with the ancilla
qubits in the bottom row. Logical qubits are formed by pairs of top row and
ancilla qubits, as indicated by the ovals. The ancilla qubits should all be
kept in the $|1\rangle $ state.}
\label{fig1}
\end{figure}

\subsection{Single-Qubit Gates}

On the logical qubits, the encoded single-qubit operations are: 
\begin{eqnarray*}
X_{m} &=&\sigma _{2m-1}^{x} \\
Z_{m} &=&\sigma _{2m-1}^{z}\otimes \sigma _{2m}^{z}.
\end{eqnarray*}
$X_{m}$ acts as a logical $\sigma ^{x}$ operation on the encoded
qubit, since: 
\[
X_{m}|0_{L}\rangle _{m}=\sigma _{2m-1}^{x}|0_{2m-1}1_{2m}\rangle
=|1_{2m-1}1_{2m}\rangle =|1_{L}\rangle _{m}.
\]
Similarly, it is simple to check that $Z_{m}$ acts as an encoded phase shift. Note
that whenever $X_{m}$ is on $Z_{m}$ is off, and {\it vice versa}, so that we can easily
implement an Euler angle rotation [Eq.~(\ref{eq:Euler}) below] to generate the Lie group $SU(2)$ on the $
m^{{\rm th}}$ logical qubit.

\subsection{Two-Qubit Gates}

It can be shown that the present model is not universal with
nearest-neighbor interactions only in 1D. Therefore we consider a quasi-1D
triangular arrangement, as shown in Figure~\ref{fig1}. Each dot represents a physical
qubit and the numbers are our indexing scheme:\ odd-numbered qubits are in
the top row. In this arrangement qubits $2m-1$, $2m$, $2m+1$ are all
nearest-neighbors and are connected by SSETs. The logical qubits are
represented by pairs, as indicated by the ovals in Figure~\ref{fig1}. The lower line
of physical qubits are all in the state $|1\rangle $. We refer to these as
the ancilla qubits.

As detailed in Ref.~\cite{Blais:00}, interaction between physical qubits is
supplied by the SSETs which couples pairs of qubits through a term $
J_{ij}\sigma _{i}^{z}\otimes \sigma _{j}^{z}$. With the above encoding, an
encoded controlled-phase gate ({\sc CPHASE}) between the $m^{{\rm th}}$ and $
(m+1)^{{\rm th}}$ logical qubits corresponds simply to coupling the
odd-numbered (top-row) physical qubits of those two logical qubits. To
see this, note that the Hamiltonian generating the {\sc CPHASE} gate is
\begin{eqnarray}
H_{m,m+1}^{\text{{\sc CPHASE}}} &=& J_{2m-1,2m+1} Z_m \otimes Z_{m+1}
\nonumber \\
&=& J_{2m-1,2m+1}\sigma _{2m-1}^{z}\otimes \sigma _{2m+1}^{z} .
\end{eqnarray}
Hence the implementation of the {\sc CPHASE} gate only involves a (quasi
1D-)nearest-neighbor two-body term, and the control of the single Josephson
energy $J_{2m-1,2m+1}$. Only odd-numbered physical qubits are involved in
this gate.

We now have all the ingredients for universal QC. However,
there are some subtleties and potential simplifications, to which we
now turn.

\section{Alternative Code Using a Bus Qubit}

\begin{figure}
\centering
\includegraphics[height=10cm,angle=270]{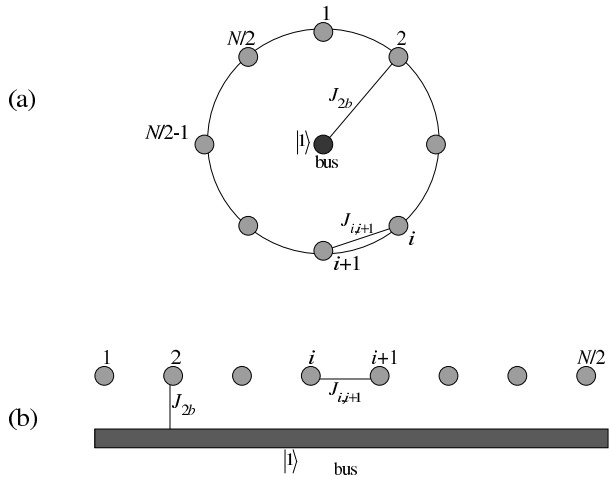}
\vspace{-1cm}
\caption{Architecture with single bus qubit. Two possible
arrangements are shown, with bus qubit at center of circularly arranged
qubits (a), or ``long'' bus qubit connected to all other qubits (b). The bus
qubit should always be in the $|1\rangle $ state.}
\label{fig4}
\end{figure}

Since the even-numbered qubits in Figure~\ref{fig1} must be kept in an identical
state there is the possibility of simply replacing them all by a {\em single}
``bus'' qubit which is kept in the $|1\rangle $ state. Thus instead of using 
$N$ physical qubits we would use only $N/2+1$, as depicted in two possible
geometries in Figure~\ref{fig4}. It is simple to check that doing so is consistent
with our logical operations:\ Since $X_{m}=\sigma _{2m-1}^{x}$ and $
H_{m,m+1}^{\text{{\sc CPHASE}}}=J_{2m-1,2m+1}\sigma _{2m-1}^{z}\otimes \sigma
_{2m+1}^{z}$, the only operation that involves the even-numbered qubits is $
Z_{m}=\sigma _{2m-1}^{z}\otimes \sigma _{2m}^{z}$. If we replace the $(2m)^{
{\rm th}}$ qubit by a fixed bus qubit $|1\rangle _{b}$ then the only change
necessary is $Z_{m}=\sigma _{2m-1}^{z}\otimes \sigma _{b}^{z}$. I.e., we
need to be able to turn on/off a Josephson coupling between all odd-numbered
qubits and the bus qubit. We remark that the situation where a qubit
in a code must be kept fixed has also arisen in the context of encoded
universality for the XY model, namely the ``truncated qubit'' of Ref.~\cite{Kempe:01a}.

\section{Problems due to Tunneling on Passive Qubits and Encoded Selective
Recoupling}

\label{problems}

There is a problem with performing universal QC using only $H_{X}+H_{ZZ}$
and the above encoding, which must now be addressed. As stressed
above, in the idle state each qubit undergoes 
Rabi oscillations. We therefore need to ``freeze'' the evolution of passive
qubits while logical operations are performed on active qubits. In Ref.~\cite{Blais:00} this was done by using recoupling pulses, a technique
similar to that used routinely in NMR \cite{Waugh:68,Slichter:book}. However, this was
implemented by single-qubit bias pulses, which, as stressed above, we do not assume are available here. To solve this problem, we therefore extend the
scheme of Ref.~\cite{Blais:00} to the technique of {\em encoded selective
recoupling} developed in Ref.~\cite{LidarWu:01}.

Like selective recoupling, encoded selective recoupling is based on some
simple identities and the elementary theory of angular momentum. Let
$A$ and $B$ be arbitrary Hermitian operators, $\theta,\phi$ real numbers. Then
\begin{equation}
e^{-i\varphi A} e^{i\theta B} e^{i\varphi A} = e^{i \theta
\exp (-i\varphi A) B \exp (i\varphi A)} .
\label{eq:ABA}
\end{equation}
Now assume $A = \hat{n}\cdot \vec{J}$ where $\hat{n}$ is a real 3D unit
vector and $\vec{J} = (J_x,J_y,J_z)$ is an angular momentum vector
operator. Then $R \equiv \exp (-i\varphi A)$ is a rotation operator that can
be written as a product of three Euler angle rotations
\cite[Sec.~13]{Rose:book}:
\begin{equation}
R = e^{-i \alpha J_z} e^{-i \beta J_y} e^{-i \gamma J_z} ,
\label{eq:Euler}
\end{equation}
where $\alpha,\beta,\gamma$ are the Euler angles.
If we take
$B$ to be an irreducible tensor operator of rank $L$ (i.e., a member
of a set of
$2L+1$ functions $T_{LM}$, $M=-L,-L+1,...,L$, which transform under
the $2L+1$-dimensional representation of the rotation group), then \cite[Sec.~17]{Rose:book}:
\[
R \, T_{LM} R^\dagger = \sum_{M'} D^L_{M'M}(\alpha,\beta,\gamma) T_{LM'} ,
\]
where $D$ are the Wigner rotation
matrices, whose matrix elements $D^L_{M'M}(\alpha,\beta,\gamma)$ are
the matrix elements of $R$ in the $LM$ representation. The Wigner
matrices are extensively tabulated \cite{Rose:book}, so that in
principle calculating the transformation $\exp (-i\varphi A) B \exp
(i\varphi A)$ appearing in Eq.~(\ref{eq:ABA}) is always possible for
angular momentum operators $A,B$. Let
us consider a few cases of importance to us here, and of general
interest in quantum computing.

Assume that $A$ and $B$ are anticommuting operators, e.g., different
Pauli matrices. Let $I$ be the identity operator. Then, we define the operation of
``conjugation by $A$'', i.e., Eq.~(\ref{eq:ABA}), as \cite{Lidar:00b}: 
\begin{eqnarray}
C_{A}^{\varphi }\circ e^{i\theta B} &\equiv & e^{-i\varphi A} e^{i\theta B} e^{i\varphi A} \nonumber \\
&=& \exp[i\theta (B\cos(2\varphi) +  iBA \sin(2\varphi))] \nonumber \\
&=&\left\{ 
\begin{array}{r}
e^{-i\theta B}\quad {\rm if\,\,}\varphi = \pm \pi /2 \\ 
e^{\pm i\theta (iBA)}\quad {\rm if\,\,}\varphi = \pm \pi /4 \\
e^{i\theta B(I \pm iA)/\sqrt{2}}\quad {\rm if\,\,}\varphi = \pm \pi /8
\end{array}
\right. .  \label{eq:C}
\end{eqnarray}
To prove these relations one can use the Wigner $D$ matrices as above,
or, more simply but less generally, note that for anticommuting $A$
and $B$ that in addition satisfy $A^2=I$:

\begin{eqnarray*}
e^{-i\varphi A} B e^{i\varphi A} &=& (I\cos \varphi -A i \sin
\varphi) B (I\cos \varphi +Ai\sin \varphi )  \nonumber \\
&=& B \cos ^{2} \varphi + A B A \sin ^{2}\varphi - i\sin \varphi \cos
\varphi [A,B] \\
&=& B\cos(2\varphi) +  i B A \sin(2\varphi) .
\end{eqnarray*}
The result of Eq.~(\ref{eq:C})\ with $\varphi =\pi /2$ is used in the NMR
technique of refocusing, or more generally, selective recoupling
\cite{Waugh:68,Slichter:book}. It allows 
one to flip the sign of a term in a Hamiltonian, which can be used to cancel
unwanted evolution. The result with $\varphi =\pi /4$ can be viewed as a
special case of Euler angle rotations which preserve a discrete group
(in QC commonly the Pauli group -- the group of tensor products of Pauli
matrices). The case with $\varphi =\pi /8$
allows us to move from the Pauli group to the group algebra of the
Pauli group, and is useful for rotating sums of operators into a
desired direction on the Bloch sphere (see Section~\ref{partsym} below). The conjugation method can be used 
on the physical as well as the encoded qubits, in which case we refer
to it as ``{\em encoded} selective recoupling''.

Note that in order to implement $e^{-itA}$ where $A$ is a Hamiltonian that
is turned on for a time $t$, we need to find $\vartheta $ such that $
e^{i\vartheta A}=I$ and implement $e^{i(\vartheta -t)A}$ instead. This
depends on $A$ having rationally related eigenvalues, which is the case
for the Hamiltonians of interest to us. E.g., for $A=J\sigma _{1}^{z}\otimes
\sigma _{2}^{z}$ we have $\vartheta =2\pi /J$, so that $\exp (-it(-J)\sigma
_{1}^{z}\otimes \sigma _{2}^{z})=\exp (i(t-2\pi /J)J\sigma _{1}^{z}\otimes
\sigma _{2}^{z})$. I.e., if this Hamiltonian is applied for a time $t-2\pi
/J>0$ it effectively evolves as if $J\rightarrow -J$. This method
circumvents the need for switching the sign of the Hamiltonian itself, which
is difficult to realize in the case of the Josephson coupling $J_{ij}$.

We are now ready to address and solve the issue of the idle qubit Rabi
oscillations mentioned above. We will
first concentrate on the realization of Figure~\ref{fig1} before turning to the bus
qubit implementation.

\subsection{Leakage}

\label{leakage}

In the idle state there will be tunneling on all qubits implementing
unwanted $\sigma ^{x}$ operations. On the even-numbered qubits of Figure~\ref{fig1}, these bit
flips cause a transition out of the computational space, or {\em leakage}.

Our goal is then to eliminate the leakage term on all even-numbered qubits 
\[
\Delta =\sum_{m=1}^{N/2}\Delta _{2m}\sigma _{2m}^{x} 
\]
due to the free evolution of these qubits during an idle period. For
simplicity consider just the first and second encoded qubits, i.e., the term 
$\Delta _{2}\sigma _{2}^{x}+\Delta _{4}\sigma _{4}^{x}$. We can refocus it
using $\sigma _{2}^{z}\otimes \sigma _{4}^{z}$, which will flip the sign of
the offending term. To do so we need to turn on $\sigma _{2}^{z}\otimes \sigma
_{4}^{z}$ for a time $\tau _{24}$ such that $\tau _{24}J_{24}=\pm \pi
/2$ [recall Eq.~(\ref{eq:C})].
Specifically, since $A=\sigma _{2}^{z}\otimes \sigma _{4}^{z}$ and $B=\Delta
_{2}\sigma _{2}^{x}+\Delta _{4}\sigma _{4}^{x} $ are Hermitian and anticommute:
\[
e^{it(\Delta _{2}\sigma _{2}^{x}+\Delta _{4}\sigma _{4}^{x})} \left(
C_{\sigma _{2}^{z}\otimes \sigma _{4}^{z}}^{\pi /2}\circ e^{it(\Delta
_{2}\sigma _{2}^{x}+\Delta _{4}\sigma _{4}^{x})}\right) = I, 
\]
so that evolution under $\Delta _{2}\sigma _{2}^{x}+\Delta _{4}\sigma
_{4}^{x}$ has been eliminated.

Note that Eq.~(\ref{eq:C}) implicitly assumes that while $A$ is on $B$ is
off, and {\it vice versa}. This is satisfied in our procedure, since the tunneling term $
\Delta _{2}\sigma _{2}^{x}+\Delta _{4}\sigma _{4}^{x}$ is on (off) while the
coupling term $\sigma _{2}^{z}\otimes \sigma _{4}^{z}$ is off
(on). Interestingly, this is different from most other recoupling
schemes, e.g., \cite{LidarWu:01,Viola:01a}, where
one typically assumes strong pulses that completely dominate the
natural evolution (see \cite{Protopopescu:02} for an analysis of this
issue). Continuing, $\Delta $ can be completely eliminated if all pairs of qubits $
\{(2,4),(6,8),...\}$ are refocused in a similar manner, i.e., by turning on $
\sigma _{4m-2}^{z}\otimes \sigma _{4m}^{z}$ for a time $\tau _{4m-2,4m}$
such that $\tau _{4m-2,4m}J_{4m-2,4m}=\pm \pi /2$ (where $m=1,2,...,N/4$).
With $J_{ij}$ of the order of the GHz, the refocusing pulses will typically be
of the order of the nanosecond. These operations need to be applied in
parallel, which is possible because they commute (they are applied to
different qubits). This means that all $J_{4m-2,4m}$ must be rationally related,
so that appropriate intervals $\tau_{4m-2,4m}$ can be found. Obviously, a
possibility is to set all $J_{4m-2,4m}$ equal but this could be difficult to realize
in practice.

\subsection{Unwanted $\protect\sigma ^{x}$ Operations During Encoded Single
Qubit Operation}

\label{unwantedX}

We now take care not only of leakage out of the code subspace, but of errors
due to tunneling on passive qubits during operation of encoded single qubit
gates on active qubits.

For example, while $X_{l}=\sigma _{2l-1}^{x}$ is on we still have tunneling
on all qubits other than $2l-1$, again implementing unwanted $\sigma ^{x}$
operations. To emphasize this let us rewrite $H_{X}$ as: 
\begin{eqnarray*}
H_{X} &=&H_{0}+H_{\bar{X}}+\Delta  \nonumber \\
&\equiv & \Delta_{2l-1}X_{l} + \sum_{m\neq l}^{N/2}\Delta
_{2m-1}X_{m}+\sum_{m=1}^{N/2}\Delta_{2m}\sigma _{2m}^{x}
\end{eqnarray*}
where $H_{0}$ is the desired evolution, $H_{\bar{X}}$ are unwanted bit
flips on the passive qubits and $\Delta $ is a leakage term.

Similarly, while $Z_{l}$ is on, we have tunneling in all qubits other than $
2l-1$ and $2l$, again implementing unwanted bit
flips. I.e.,
the Hamiltonian while $Z_{l}$ is on is: 
\begin{equation}
H_{Z_{l}}=J_{l}Z_{l}+\sum_{m\neq l}^{N/2}\Delta
_{2m-1}X_{m}+\sum_{m\neq l
}^{N/2}\Delta _{2m}\sigma _{2m}^{x}  \label{eq:Zl}
\end{equation}
Thus we either have to implement computational operations on the other
encoded qubits ($m\neq l$), or eliminate this unwanted evolution.

\begin{figure}
\centering
\includegraphics[height=10cm,angle=270]{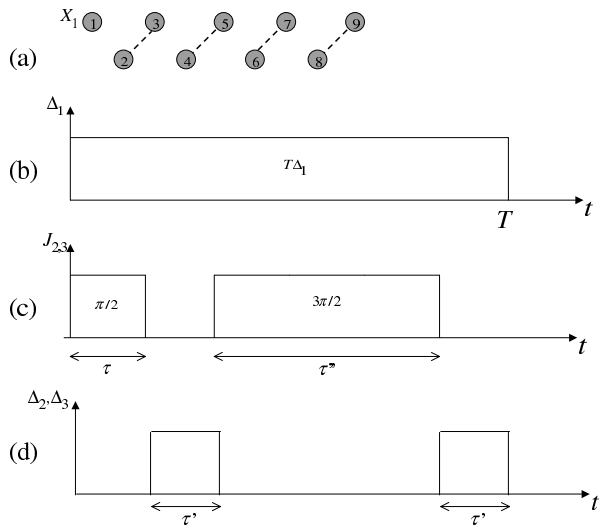}
\caption{Single logical-qubit $X$ operation. Shown in (a) is the
algorithm for the implementation of $X_{1}$, with recoupling connections
indicated by dashed lines. (b) Shows a rectangular pulse with an area $
T\Delta _{1}$ implementing the logical operation $\exp (-iT\Delta _{1}X_{1})$. (c),(d) Show the corresponding recoupling sequence on qubits $2,3$
(exactly the same diagrams apply to the other qubit pairs $4,5$ etc.). E.g.,
in (c) the first pulse lasts for a time $\tau $ such that $\tau J_{2,3}=\pi
/2$, immediately followed by a tunneling period $\tau ^{\prime }$ (assuming $
\Delta _{2}=\Delta _{3}$), etc.}
\label{fig2}
\end{figure}

We start by first taking care of the single encoded $X_{l}=\sigma _{2l-1}^{x}
$ operation in the presence of tunneling on all other qubits. I.e., we need
to eliminate not just the leakage term $\Delta $ but also the unwanted
logical operations $H_{\bar{X}}$. Suppose we wish to implement $X_{1}$ for a
time $T$, i.e., the operation $ e^{-iT\Delta _{1}X_{1}}$. The solution is
shown schematically in Figure~\ref{fig2}. We refocus all other qubits in pairs $
\{(m,n)\}= \{(2,3),(4,5),(6,7),(8,9),...\}$, using the same idea as for the
leakage problem. Namely, we turn on the interactions $\{\sigma
_{m}^{z}\otimes \sigma _{n}^{z}\}$ for times $\tau _{mn},\tau _{mn}^{\prime
\prime }$ such that $\tau _{mn}J_{mn}=\pi /2$ and $\tau _{mn}^{\prime \prime
}J_{mn}=3\pi /2 $ (we are now explicitly taking into account the fact that
we cannot switch the sign of $J_{mn}$). In between and after the $\pi /2$
and $3\pi /2$ periods these pairs of qubits are allowed to evolve freely for
times $\tau _{mn}^{\prime }$ under the tunneling terms $\{\Delta _{m},\Delta
_{n}\}$. The condition that must then be satisfied is: 
\[
\tau _{mn}+2\tau _{mn}^{\prime }+\tau _{mn}^{\prime \prime }=T. 
\]
Since, e.g., $\tau _{23}J_{23}=\pi /2$, the free evolution time is
determined by $\tau _{23}^{\prime }=\left( T-4\tau _{23}\right) /2=T/2-\pi
/J_{23}$, or, in general \footnote{This is for the case where $J_{23}$
is kept fixed over the periods $\tau$ and
$\tau''$.  One could also adjust the strength of $J_{23}$ (voltage on
the SSET) to have, e.g.,
 $\tau=\tau''$. Which adjustment to make will be a matter of
experimental convenience.}: 
\[
\tau _{mn}^{\prime }=T/2-\pi /J_{mn}. 
\]
These times must be positive, so that we must be able to make $J_{mn}$
(assuming it is positive) large enough that 
\[
J_{mn}>2\pi /T. 
\]
Conversely, if $J_{mn}$ has a maximum value $J_{mn}^{\max }$ then there is a
time 
\[
T_{\min }=2\pi /J_{mn}^{\max } 
\]
below which we cannot apply an encoded logical $X$ operation. Thus all encoded logical $X$ operations must be
implemented as $e^{-iT\Delta _{1}X_{1}}$ with $T\geq T_{\min }$. This means
that the smallest angle of rotation around the encoded $x$ axis is 
$\theta_{\min }=2\pi \Delta_{m(n)} /J_{mn}^{\max }$.  With $\Delta_{m}\sim 100$MHz
\cite{Zagoskin:99} and $J_{mn}$ of the order of the GHz, $\theta_{\min}$ is of
the order of $2\pi/10$ and can be made smaller if the Josephson energy
$J_{mn}$ is made larger. However, this is not necessary in the
fault-tolerant ``standard paradigm'' of universal QC, where all logic
gates can be built up from the Hadamard, $\pi/8$ and {\sc CPHASE}
gates, which require $\theta$ of $\pi/2$, $\pi/8$, and $\pi/4$
respectively \cite{Boykin:99}. Of course, here we have in mind an {\em
encoded} version of this universal set of gates.

\begin{figure}
\centering
\includegraphics[height=10cm,angle=270]{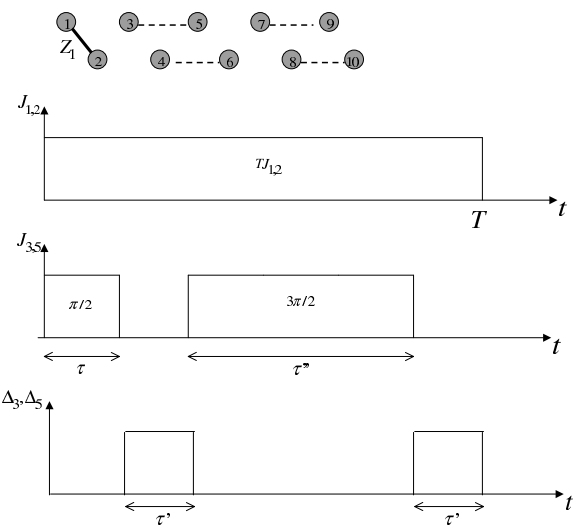}
\caption{Single logical-qubit $Z$ operation. Other details as in
Figure~\ref{fig2}.}
\label{fig3}
\end{figure}

The solution to the single encoded $Z_{l}=\sigma_{2l-1}^{z}\otimes \sigma
_{2l}^{z}$ operation in the presence of tunneling on all other qubits, 
Eq.~(\ref{eq:Zl}), is almost identical to that of single encoded $X$ operations.
Suppose we wish to implement $Z_{1}=\sigma _{1}^{z}\otimes \sigma _{2}^{z}$
for a time $T$, i.e., the operation $e^{-iTJ_{1}Z_{1}}$. The solution is
shown schematically in Figure~\ref{fig3}. We refocus all other qubits in pairs $
\{(m,n)\}=\{(3,5),(4,6),(7,9),(8,10),...\}$ by again turning on the
interactions $\{\sigma _{m}^{z}\otimes \sigma _{n}^{z}\}$ for times $\tau
_{mn},\tau _{mn}^{\prime \prime }$ such that $\tau _{mn}J_{mn}=\pi /2$ and $
\tau _{mn}^{\prime \prime }J_{mn}=3\pi /2$. All other details are the same
as above.

\subsection{Unwanted $\protect\sigma ^{x}$ Operations During Two Qubit
Encoded $Z\otimes Z$ Operation}

\label{unwantedZZ}

The problem is the same as in the previous cases. We now have to deal with
the Hamiltonian 
\begin{equation}
H_{Z_{l}Z_{l+1}}=J_{2l-1,2l+1}Z_{l}Z_{l+1}+\sum_{m\neq
l,l+1}^{N/2}\Delta_{2m-1}X_{m}+\Delta ,  \label{eq:ZlZl+1}
\end{equation}
i.e., leakage plus unwanted $X$ operations on all but encoded qubits
$l,l+1$. The solution is completely analogous to the case of single
encoded $Z$ operations.

\subsection{Compatibility of Recoupling Sequences}

\label{compatibility}

\begin{figure}
\centering
\includegraphics[height=10cm,angle=270]{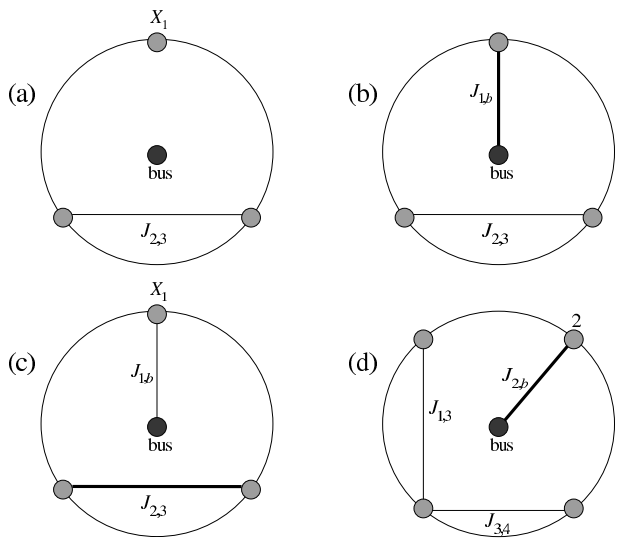}
\caption{Logic operations with recoupling connections in bus qubit
architecture. Recoupling works well with an odd number of non-bus qubits, as
shown in (a)-(c). (a) Single logical-qubit $X_{1}$ operation, qubits $2,3$
recoupled. Here it is essential that the bus qubit can be kept fixed in the $
|1\rangle $ state by a method other than recoupling; (b) Single
logical-qubit $Z_{1}$ operation, qubits $2,3$ recoupled; (c) Two
logical-qubit {\sc CPHASE} operation, qubit $1$ recouples with bus qubit.
With an even number of non-bus qubits there is a problem, as shown in (d):
it is not possible to implement a single logical-qubit $Z_{2}$ operation
because three non-bus qubits have to be recoupled.}
\label{fig5}
\end{figure}

There is still one problem with the above method that needs to be solved: recoupling in the single encoded $X$ case requires pairing up $N-1$
physical qubits, whereas recoupling in the single encoded $Z$ and encoded $
Z\otimes Z$ cases requires pairing up $N-2$ physical qubits. Thus in the
former case we would want $N$ to be odd, while in the latter we need $N$ to
be even. Assuming $N$ to be even, there are several possible solutions to this.

One solution is to reintroduce single-qubit $\sigma^z$ operations, but only
on (say) the last qubit. This qubit can then be refocused independently.
This is however not very elegant (or practical) as our aim was to eliminate single-qubit 
$\sigma^z$ operations in the first place.

A second possibility is to make the Josephson coupling energy much larger
than the tunneling matrix element. In this case, it is possible to neglect
the evolution of the passive qubits due to tunneling during the on-time of
the Josephson coupling on the active qubit. Reaching the necessary
constraint $|J_{ij}|\gg |\Delta _{k}| \; \forall k$ can however be
difficult in practice and, even if it is reached, there will be accumulation
of small errors on the passive qubits. This solution is therefore not the
most practical either.

The simplest way to solve the compatibility problem is to assume that we can
fix (say) the last qubit (number $N$) in the $|1\rangle $ state. This
corresponds to producing that qubit with a very high tunneling barrier and
initializing it in the $|1\rangle $ state. In this manner this qubit is
``frozen'' and the problem of unwanted evolution on it does not arise. For the
dGB qubit, producing a qubit with a high tunneling barrier
can be done by choosing proper misalignment between the two d-wave
superconductors or, more simply, by working with larger junctions 
\cite{Zagoskin:99}.

An interesting alternative is to extend this concept of ``frozen'' qubit to all
ancilla qubits. In this case, there is no possible leakage out of the code
subspace. Moreover, this applies well to the concept of ``bus qubit''
presented above. In this case, the bus is a single qubit with a large
tunneling barrier and initialized to the $| 1 \rangle$ state.

Considering the bus qubit case, even if there is no leakage, the problem of freezing passive qubits is still
present. Fortunately, the solution explained above still applies. 
However, as illustrated in Figure~\ref{fig5}, with a bus qubit the number of physical
qubits should be odd. The reason is that otherwise, during the
implementation of, e.g., $Z_{1}=\sigma _{1}^{z}\otimes\sigma _{b}^{z}$ we
would be left with an odd number of qubits to be refocused [Figure~\ref{fig5}(d)].
As seen in Figure~\ref{fig5}((a)-(c)), with an odd number of logical qubits all
logic operations can be performed provided the bus qubit can be kept fixed
in the $|1\rangle $ state during single qubit $X$ operations [Figure~\ref{fig5}(a)].

The question of which of the two designs, bus qubit or with $N/2$ ancilla
qubits, is superior, will be decided by engineering constraints. The bus
qubit imposes a circular geometry [as in Figure~\ref{fig4}(a)], or has to be made
long [as in Figure~\ref{fig4}(b)], or some other method has to be found to connect it
to all other qubits. In particular, it can be challenging experimentally to
connect many qubits to the bus. Having several bus qubits and, for example,
repeating the circular geometry of Figure~\ref{fig4}(a) in an triangular lattice could
help in reducing these constraints.

\section{Codes with Higher Rates}

So far we have considered encoding a single logical qubit into two
physical qubits. This code has a rate of $1/2$. In this section we
consider codes with higher rates. Specifically, we propose an encoding
of two logical qubits into three physical qubits, yielding a rate of
$2/3$.

\begin{figure}
\centering
\includegraphics[height=10cm,angle=270]{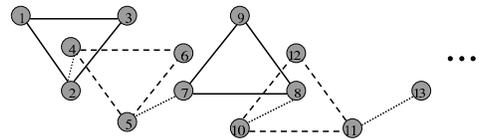}
\vspace{-5cm}
\caption{3D architecture using 3 physical qubits to encode 2 logical
qubits. Two triangular 2D arrays are superimposed (alternating solid
and dashed lines), such that qubits 1,2,3,4 form a tetrahedron,
likewise for qubits 3,6,7,9, and 8,10,11,12, etc. Pairs of logical
qubits correspond to physical qubits 123, 456, 789, etc. This allows
qubits 12,23,13,24 to have nearest neighbor interactions, so that the
logical operations $Z_1$, $Z_2$, $Z_1 Z_2$, $Z_2 Z_3$
can all be implemented through nearest neighbor couplings. Likewise,
qubits 45,56,46,57 are nearest neighbors, allowing logical operations
$Z_3$, $Z_4$, $Z_3 Z_4$, $Z_4 Z_5$ to be implemented
through nearest neighbor couplings. The required connections between
qubits belonging to different planes are shown by short-dashed lines.}
\label{fig2in3}
\end{figure}

Consider the following encoding:
\begin{eqnarray*}
|0_{L}\rangle\otimes|0_{L}\rangle &  =& |0_{1}0_{2}0_{3}\rangle\\
|0_{L}1_{L}\rangle &  =& |010\rangle\\
|1_{L}0_{L}\rangle &  =& |100\rangle\\
|1_{L}1_{L}\rangle &  =& |110\rangle.
\end{eqnarray*}
The third physical qubit is always $0$ here, so once again the bus
qubit idea applies. The corresponding logical operations are:
\begin{eqnarray*}
X_{1}  &=& \sigma_{1}^{x} \quad X_{2}  = \sigma_{2}^{x}\\
Z_{1}  &=& \sigma_{1}^{z}\sigma_{3}^{z} \quad Z_{2}  = \sigma_{2}^{z}\sigma_{3}^{z}\\
Z_{1}Z_{2}  &=& \sigma_{1}^{z}\sigma_{2}^{z} ,
\end{eqnarray*}
as is readily checked. E.g.,
\begin{eqnarray*}
Z_{1}Z_{2}|0_{L}0_{L}\rangle &=& \sigma_{1}^{z}\sigma_{2}^{z}|000\rangle
=|000\rangle=|0_{L}0_{L}\rangle\\
Z_{1}Z_{2}|0_{L}1_{L}\rangle &=& \sigma_{1}^{z}\sigma_{2}^{z}|010\rangle
=-|010\rangle=-|0_{L}1_{L}\rangle,
\end{eqnarray*}
etc. Note that $Z_{1}$ and $Z_{2}$ explicitly use the third qubit, showing that
it is essential. As before, we think of these logical operations as
Hamiltonians, not gates. We therefore have a universal generating set for two logical qubits encoded
into the first three physical qubits. To complete the universal set we also
need to be able to couple logical qubits belonging to different blocks of
three physical qubits. Let physical qubits $4,5,6$ encode the next pair of
logical qubits. Then
\[
Z_{2}Z_{3}=\sigma_{2}^{z}\sigma_{4}^{z}
\]
is a logical operation coupling the two blocks. To verify this consider
the truth table for this operation. Let $x,y$ be $0$ or $1$. Then:
\begin{widetext}

\begin{eqnarray*}
Z_{2}Z_{3}|x_{L}0_{L}\rangle|0_{L}y_{L}\rangle &=& \sigma_{2}^{z}\sigma
_{4}^{z}|x_{1}0_{2}0_{3}\rangle|0_{4}y_{5}0_{6}\rangle=|x_{1}0_{2}0_{3}
\rangle|0_{4}y_{5}0_{6}\rangle=|x_{L}0_{L}\rangle|0_{L}y_{L}\rangle\\
Z_{2}Z_{3}|x_{L}0_{L}\rangle|1_{L}y_{L}\rangle &=& \sigma_{2}^{z}\sigma
_{4}^{z}|x_{1}0_{2}0_{3}\rangle|1_{4}y_{5}0_{6}\rangle=-|x_{1}0_{2}
0_{3}\rangle|1_{4}y_{5}0_{6}\rangle=-|x_{L}0_{L}\rangle|1_{L}y_{L}\rangle\\
Z_{2}Z_{3}|x_{L}1_{L}\rangle|0_{L}y_{L}\rangle &=& \sigma_{2}^{z}\sigma
_{4}^{z}|x_{1}1_{2}0_{3}\rangle|0_{4}y_{5}0_{6}\rangle=-|x_{1}1_{2}
0_{3}\rangle|0_{4}y_{5}0_{6}\rangle=-|x_{L}1_{L}\rangle|0_{L}y_{L}\rangle\\
Z_{2}Z_{3}|x_{L}1_{L}\rangle|1_{L}y_{L}\rangle &=& \sigma_{2}^{z}\sigma
_{4}^{z}|x_{1}1_{2}0_{3}\rangle|1_{4}y_{5}0_{6}\rangle=|x_{1}1_{2}0_{3}
\rangle|1_{4}y_{5}0_{6}\rangle=|x_{L}1_{L}\rangle|1_{L}y_{L}\rangle ,
\end{eqnarray*}

\end{widetext}
which is the desired action. Note that next-nearest neighbor couplings are involved in $Z_{1}$ (inside a
block of three) and $Z_{2}Z_{3}$ (between blocks of three).  Figure~\ref{fig2in3} shows a possible arrangement
in 3D. Two triangular arrays of qubits are superimposed such that the relevant
qubits form tetrahedrons (see caption).  With this particular geometry, only nearest neighbor couplings are required.
Qubit arrays of lower dimensionality will require next-nearest
neighbor couplings, which may or may not be easier to implement
experimentally. For definiteness we continue the discussion here while
referring to the 3D layout, but it should be kept in mind that a 1D/2D
layout with overlapping wires may prove to be advantageous.

\begin{figure}
\centering
\includegraphics[height=10cm,angle=270]{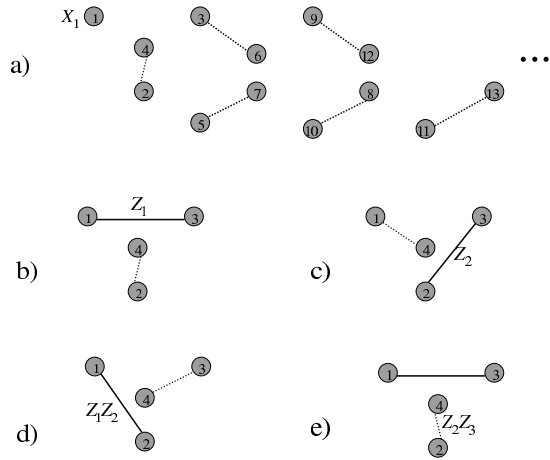}
\vspace{-1cm}
\caption{(a) Implementation of logical $X_1$ operation on logical qubit 1 is
done by allowing physical qubit 1 to undergo free tunneling. This
requires recoupling all other physical qubits in pairs as indicated by
short-dashed lines. 
(b) Implementation of logical $Z_1$ operation on logical qubit 1 is done by
Josephson coupling physical qubit 1 to its ancilla qubit, number 3. This
requires recoupling all other physical qubits in pairs as indicated by
dashed lines.
(c),(d),(e) Same as (b), for implementing logical operations $Z_2$, $Z_1
\otimes Z_2$, $Z_2 \otimes Z_3$.}
\label{fig2in3X}
\end{figure}

Recoupling will still be needed since we still have passive tunneling on the
idle physical qubits. The recoupling paths are shown in
Figure~\ref{fig2in3X}. As in the rate $1/2$ code, there is an
incompatibility between the number of qubits required for recoupling in the
case of a logical $X$ operation, and the number required for the logical $Z$
and $ZZ$ operations. This can be solved by introducing an auxiliary qubit. This qubit then has to be frozen in the $|0\rangle$ state,
as discussed above.

\begin{figure}
\centering
\includegraphics[height=10cm,angle=270]{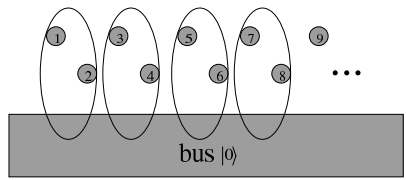}
\vspace{-5cm}
\caption{Architecture using 3 physical qubits to encode 2 logical
qubits, using a global bus qubit. Ovals indicate grouping, i.e.,
blocks of three physical qubits are formed by physical qubit 1,2,bus;
3,4,bus; etc. The bus is fixed in the $|0\rangle$ state.
}
\label{fig2in3bus}
\end{figure}

\begin{figure}
\centering
\includegraphics[height=9cm,angle=270]{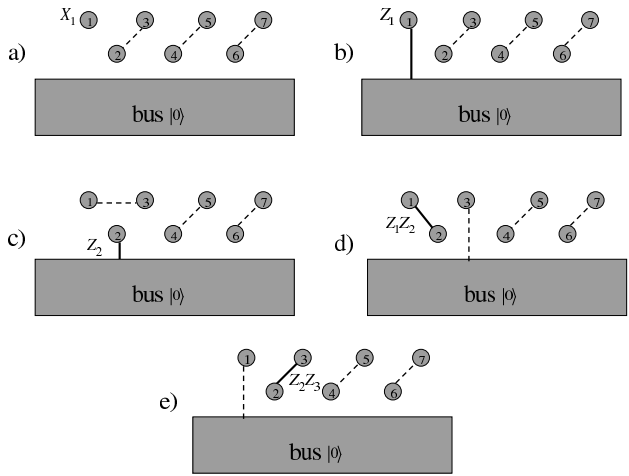}
\caption{(a) Implementation of logical $X$ operation on logical qubit 1
is done by allowing physical qubit 1 to undergo free tunneling. This
requires recoupling all other physical qubits in pairs as indicated by
dashed lines.
(b) Implementation of logical $Z$ operation on logical qubit 1 is done
by Josephson coupling physical qubit 1 to bus. This requires
recoupling all other physical qubits in pairs as indicated by dashed
lines. Similarly for (c),(d), and (e).
}
\label{fig2in3Xbus}
\end{figure}

All the considerations above can be modified easily to the case of a single
bus qubit, taken to be, e.g., a long single qubit fixed in the $|0\rangle$
state (i.e., the single bus qubit now replaces all physical qubits numbered
$3n$, $n=1,2,...$ in the discussion above). This architecture is shown in
Figure~\ref{fig2in3bus}. The recoupling paths are shown in Figure~\ref{fig2in3Xbus} (compare to the recoupling
figures for the one-in-two encoding). However, note that with a \textquotedblleft
global\textquotedblright\ bus qubit the advantage of a higher rate disappears.
As in the above encoding of one logical qubit per two physical qubits,
with a global bus qubit optimal use
of all physical qubits has already been made.

There will be codes with even higher rates (e.g., use four physical qubits to
encode three logical qubits by fixing the fourth physical qubit). However, in
this case the geometrical constraints may become impossible to
satisfy even in 3D if we can only use nearest-neighbor
coupling. Therefore such higher rate codes will require introducing
longer-range coupling. What the tradeoffs are between a 3D arrangement and
long-range couplings, is, again, an implementation-dependent question
that we cannot answer here.

\section{Implementation and connection to other designs}

We now turn to a discussion of the usefulness and practicality of the above
scheme to the dGB qubits and to the other superconducting
phase-qubits in general.

\begin{figure}
\centering
\includegraphics[height=6cm]{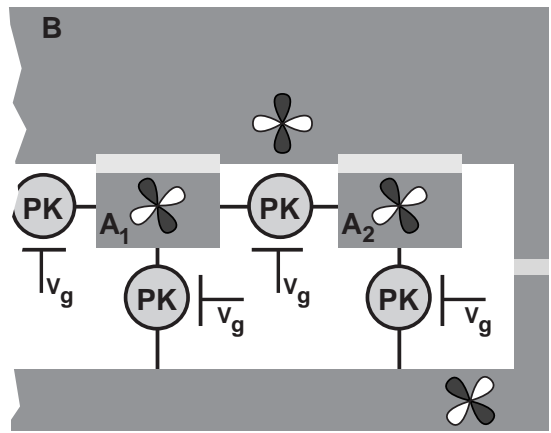}
\caption{Superconducting grain boundary qubits with ``bus qubit''. The qubits
are connected to the bus by SSETs.}
\label{fig_gbqb_bus}
\end{figure}

For the dGB qubits, the usefulness of the above scheme is
clear. The application of individual qubit bias following the scheme of Refs.~\cite{Zagoskin:99,Blais:00} would be rather difficult experimentally. The
encoding and bus qubit proposed here circumvent this problem. Moreover, for
the dGB qubits, the implementation of the bus qubit
concept turns out to be a rather simple modification of the layout already
presented in \cite{Zagoskin:99,Blais:00}. As shown in Figure~\ref{fig_gbqb_bus},
the bus can be a large piece of d-wave superconductor coupled to terminal B
of the qubits by a weak link. If this bus is large and has the proper
misalignment of its order parameter with respect to terminal B, it will have
a fixed phase and hence correspond to a fixed logical state. This bus is
coupled to the other qubits by SSETs to provide the necessary $J_{ij}$
couplings.  Thus, based on the encoding of Eq.~(\ref{eq:encoding}) which
suggested the use of a bus qubit, the equivalent of single-qubit bias
operations could be reintroduced in a way that is experimentally simpler
than what was previously suggested \cite{Zagoskin:99,Blais:00}.

\begin{figure}
\centering
\includegraphics[height=5.5cm]{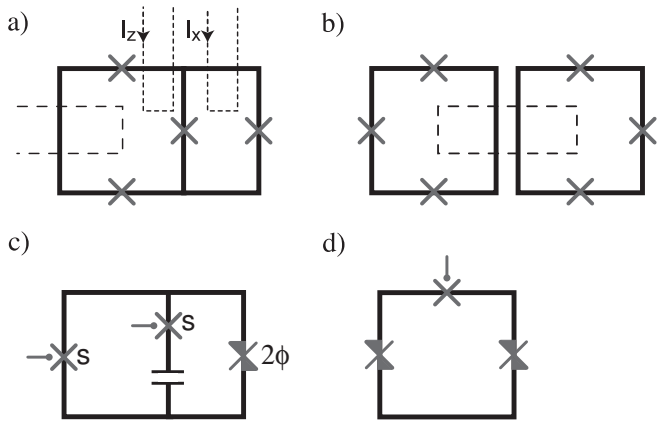}
\caption{(a) Original persistent-current qubit design.  Currents $I_z$
and $I_x$ provide flux in large and small loop respectively and are
used to control single-qubit operations. Coupling to a second qubit
(not shown) can be inductive (long-dashed line). (b) Using the
encoding of Eq.~(\ref{eq:encoding}), a pair of coupled physical qubits
can be used as one logical qubit.  Unnecessary control lines have been
discarded, keeping only the inductive two-qubit coupling. (c) `Quiet
qubit' design. The $2\phi$-junction realizes the two-level
system. Single qubit operations are implemented by voltage pulses on
the switches `s'. Coupling to other qubits is not shown.  (d) As in
(b), only coupling between a pair of physical qubits is retained to
obtain one logical qubit.}
\label{fig_squids}
\end{figure}

The concepts explored in this paper can be applied to the other
superconducting phase qubit designs \cite{Blatter:01,Mooij:99,Orlando:99,Amin:01}. 
First, for the rf-SQUID like
design of Ref.~\cite{Mooij:99,Orlando:99}, single-qubit logical operations
are provided by control of external fluxes in two different loops of the
SQUID, Figure~\ref{fig_squids}a). One of these fluxes controls the asymmetry of the qubit's potential
energy landscape (i.e., $\sigma_i^z$ operations), while the other controls the
tunnel barrier (i.e., $\sigma_i^x$ operations). Qubits can be coupled
inductively to provide the necessary two-qubit logical gates. These coupling
can be arranged to provide a term in the Hamiltonian which has the same
symmetry as considered above, $J_{ij}\,\sigma_i^z\otimes\sigma_j^z$
\cite{Orlando:99}.

Consider moving to a design with only 3 junctions and one loop (as was used
experimentally in \cite{vdWal:00}), such that only the asymmetry can be
controlled, Figure~\ref{fig_squids}b). By choosing the Josephson energies such that in the idle
state coherent tunneling is possible, the Hamiltonian describing the system
is (\ref{gbqb_hamiltonian}), with the same constraints as for the dGB qubit. The encoding and results presented here therefore
apply equally to these qubits and provide enough control to perform
universal QC. Of course, the application of the logical
operations is more laborious than in the original 4-junction design but this
has other advantages. First, there is some simplification in fabrication.
Moreover, the external flux controlling the tunnel barrier is a source of
noise to the system. Removing such a source of decoherence in a system which
has a small coherence time to begin with, is clearly advantageous. For
this design, rather than the simplification in fabrication, the possibility
of reducing sources of noise is the real advantage of using the
encoding~(\ref{eq:encoding}).

Similarly, for the ``quiet qubit'' design of Ref.~\cite{Blatter:01}, one could
eliminate the external capacitor which is used to tune the tunneling
amplitude, and choose the Josephson energies such that tunneling is possible
in the idle state, Figure~\ref{fig_squids}c). Connection to the strong $\pi/2-$junctions, used to
control the asymmetry, can also be discarded, retaining only two-qubit
couplings, Figure~\ref{fig_squids}d). In this way the system again is described by (\ref{gbqb_hamiltonian}) and the same ideas apply. As with the rf-SQUID design,
use of the encoding (\ref{eq:encoding}) allows to remove some of the
components of the quantum computer design while maintaining
universality. We again stress that this reduces engineering constraints and
removes some important sources of noise.

Implementation of the bus qubit for the ``quiet qubit" design is similar to the phase bus discussed in Ref.~\cite{Blatter:01}.  Similar ideas can be used for the rf-SQUID like qubits.

By using the encoding suggested here, we are now relying mostly on two-qubit
gates to perform QC. Therefore, qubit-qubit coupling should
be easily controlled and the ratio of the characteristic energy when the
qubit-qubit ``switch'' is on compared to when it is off should be high. These
requirements are not different from the standard situation when
the encoding (\ref{eq:encoding}) is not used. Further, for this scheme to
be useful in practice, the two-qubit gates should not be much slower than the
1-qubit gates. By choosing strong enough Josephson junctions for the SSETs,
this should be satisfied for the dGB qubits. Similarly,
in the design of Refs.~\cite{Mooij:99,Orlando:99} the inductive coupling
qubits should be of the order of single-qubit energies. The situation is
similar for the ``quiet qubit'' design. All of the above remarks apply equally
for the multi-terminal superconducting phase qubit \cite{Amin:01}.

\section{Reduction of Errors due to Decoherence}

Standard quantum error correction techniques \cite{Steane:99} are
certainly compatible with the dGB qubit design. Here we focus on the
recoupling
\cite{Waugh:68,Slichter:book,LidarWu:01,Jones:99,Leung:00} and the
decoupling, or
``bang-bang'' (BB) method introduced in \cite{Viola:98}, and developed
further, e.g, in
Refs.~\cite{Duan:98e,Zanardi:98b,Vitali:99,Viola:99,Viola:99a,Viola:00a,Vitali:01,ByrdLidar:01,ByrdLidar:01a,WuLidar:01b,WuByrdLidar:02}.
The advantage of the recoupling and BB methods, is that they does not require extra qubit (space)
resources, unlike quantum error correction \cite{Steane:99} and/or
decoherence-free subspaces \cite{Zanardi:97c,Lidar:PRL98}. Specifically, we will show that decoherence can be reduced
significantly using only the {\em existing controllable
interactions}, acting on the encoded qubits. Different methods will be
presented depending on the symmetry
of the system-bath interaction. ``Encoded
decoupling'' (i.e., acting on encoded qubits) of the type we discuss
below has been previously suggested for solid-state
\cite{ByrdLidar:01a,WuLidar:01b,WuByrdLidar:02} and NMR
\cite{Viola:01a,Fortunato:01} QC proposals. We further note that use of BB pulses for dGB qubits was previously discussed in
Ref.~\cite{Blais:00}, but access to single-qubit $Z$ operations was
assumed. We extend and generalize this discussion here.

The total Hamiltonian of a qubit system ($S$) and bath ($B$) can be written as 
\[
H=H_{S}+H_{B}+H_{SB},
\]
where $H_{SB}$ is the system-bath coupling. In order to use the BB
method one makes two assumptions: First, (as in the
threshold result of fault tolerant QC \cite{Aharonov:96,Knill:98})
that a controllable part of $H_{S}$ can be made so
strong that we can make $H$ approximate that part of $H_{S}$ to
arbitrary precision. This is needed so that $H_{SB}$ can be neglected
during the pulse. Second, that one can
pulse this controllable part of $H_{S}$ with a pulse repetition time
that is much shorter than the inverse of the bath frequency
cutoff. This requirement arises since in between BB pulses the bath
should not change, or else it will acquire phases that, when the bath
is traced over, will result in decoherence in the system \cite{Viola:98}.

We will take the interaction term to have the linear form $H_{SB}=\sum_{i}
\vec{\sigma }_{i}\cdot \vec{B}_{i}$, where $
\vec{\sigma }_{i}$ is the vector of Pauli matrices acting on the
system, and $\vec{B}_{i}$ are corresponding bath operators. This
can represent the interaction of a qubit with a fluctuating control field.
For example, for the superconducting phase qubit, a term $\sigma_i^zB_i^z$
arises due to fluctuation of local magnetic field. The relation of $H_{SB}$ to the
parameters of the system and bath can be analyzed in detail
\cite{Makhlin:01}. Our approach to decoherence suppression depends on the
time-scales that emerge from this analysis, and on the symmetry of the system-bath
interaction.

\subsection{Suppression of Axially Symmetric System-Bath Interaction by $
Z\otimes Z$ Recoupling Method}

Suppose that $|J_{ij}|\gg |\Delta _{i}|,|B_{i}^{x}|,|B_{i}^{y}|,|B_{i}^{z}|$
so that the strong parts of $H_{S}$ are 
\[
H_{ij}=J_{ij}\sigma _{i}^{z}\otimes \sigma _{j}^{z}, 
\]
which we can turn on and off freely. If $H_{SB}$ is of the general form $
\sum_{i}\vec{\sigma }_{i}\cdot \vec{B}_{i}$, then $
H_{ij}$ obviously is not enough to eliminate $H_{SB}$ by decoupling methods,
since it commutes with the $\sigma_{i}^{z}B_{i}^{z}$ terms. However, by the
same token, if the system-bath interaction has an {\em axial symmetry} so
that 
\[
H_{SB}=\sum_{i}\left( \sigma _{i}^{x}B_{i}^{x}+\sigma
_{i}^{y}B_{i}^{y}\right) , 
\]
it can be eliminated by 
\begin{equation}
\exp (-iH_{SB}t/2)\left[ C_{\sum_{m=1}^{N/2}Z_{m}}^{\pi /2}\circ \exp
(-iH_{SB}t/2)\right] =I,
\label{eq:axsym}
\end{equation}
where as before $Z_{m}=\sigma _{2m-1}^{z}\otimes \sigma _{2m}^{z}$ is the
logical $Z$ operation. Implicit in this calculation is that $H_{SB}$ is
negligible while $Z_{m}$ is on. To the extent that this assumption breaks
down there will be an error proportional to the ratio of the largest
eigenvalue of $H_{SB}$ by the smallest eigenvalue of $H_{S}$. The important
point is that this type of axially symmetric system-bath interaction can be
suppressed {\em without any extra resources}, simply by using the already
available Josephson coupling. Moreover, the decoupling pulse
$C_{\sum_{m=1}^{N/2}Z_{m}}^{\pi /2}$ used in Eq.~(\ref{eq:axsym})
commutes with all $Z_{m'}$, $Z_{m'}Z_{m''}$ and with all $X_{m'}$ such
that $m',m'' \neq m$. All these logical operations can therefore be
executed in parallel with this decoupling procedure. However, since
the decoupling pulses anticommute with $X_m$ this logical operation
is eliminated if it is turned on during decoupling. Hence we must
alternate suppressing decoherence on the $m^{\rm th}$ qubit and
performing logical $X$ operations on it. This implies that this qubit
will suffer some decoherence  while $X_m$ is applied to it, unless we
protect it by other means, such as active quantum error correction
\cite{Steane:99}.

\subsection{Suppression of Partial Symmetric System-Bath Interaction Hamiltonians
by Decoupling}

\label{partsym}

Suppose that $|J_{ij}|\gg |\Delta _{i}|\gg
|B_{i}^{x}|,|B_{i}^{y}|,|B_{i}^{z}|$, so that we can freely turn $\Delta
_{i}\sigma^x_{i}$ and $J_{ij}\sigma _{i}^{z}\otimes \sigma _{j}^{z}$
on and off, and while we do so $H_{SB}$ becomes
negligible. The only effect of $H_{SB}$ is to decohere the qubit system when
it evolves freely, i.e., under $\exp (-it(H_{B}+H_{SB}))$. 

Now suppose the system-bath interaction has a symmetry so that 
\[
H_{SB}^{yz}=\sum_{i}\left( \sigma _{i}^{y}+\sigma _{i}^{z}\right)
B_{i}^{yz}+\sigma _{i}^{x}B_{i}^{x}. 
\]
Note that, using Eq.~(\ref{eq:C}), a term of the form $\exp i\theta (\sigma ^{y}+\sigma ^{z})$ can be
rotated to $\exp( i\theta \sigma ^{y})$:
\[
C_{\sigma ^{x}}^{-\pi /8}\circ \exp \left[ i\theta (\sigma ^{y}+\sigma
^{z})/ \sqrt{2}\right] =\exp \left( i\theta \sigma ^{y}\right) . 
\]
Hence 
\begin{eqnarray*}
C_{\sigma _{i}^{x}}^{-\pi /8}\circ \exp \left[ it\left( \left( \sigma
_{i}^{y}+\sigma _{i}^{z}\right) B_{i}^{yz}+\sigma _{i}^{x}B_{i}^{x}\right) 
\right] = \\
\exp \left[ it\left( \sqrt{2}\sigma _{i}^{y}B_{i}^{yz}+\sigma
_{i}^{x}B_{i}^{x}\right) \right] . 
\end{eqnarray*}
But this we can eliminate using the Josephson coupling, since conjugation by 
$\sigma _{i}^{z}\otimes \sigma _{i+1}^{z}$ will flip the sign of both $
\sigma _{i}^{y}$ and $\sigma _{i}^{x}$: 
\begin{eqnarray*}
\exp \left[ it\left( \sqrt{2}\sigma _{i}^{y}B_{i}^{yz}+\sigma
_{i}^{x}B_{i}^{x}\right) \right] \times \\
\left( C_{\sigma _{i}^{z} \otimes \sigma
_{i+1}^{z}}^{\pi /2}  \circ \exp \left[ it\left( \sqrt{2}\sigma
_{i}^{y}B_{i}^{yz}+\sigma _{i}^{x}B_{i}^{x}\right) \right] \right) =I. 
\end{eqnarray*}
Performing this in parallel for $i=1,3,...$ will suppress the system-bath
interaction completely. Again, there is the implicit assumption that during
the on-time of the tunneling and Josephson operations $H_{SB}$ is negligible.
Regarding computation,
similar comments as in the previous subsection apply, i.e., those logical operations that commute with the
decoupling pulses can be turned on simultaneously with the latter,
while those that anticommute cannot. 
However, the cases
\[
H_{SB}^{xy}=\sum_{i}\left( \sigma _{i}^{x}+\sigma _{i}^{y}\right)
B_{i}^{xy}+\sigma _{i}^{z}B_{i}^{z} 
\]
and
\[
H_{SB}^{zx}=\sum_{i}\left( \sigma _{i}^{z}+\sigma _{i}^{x}\right)
B_{i}^{zx}+\sigma _{i}^{y}B_{i}^{y} 
\]
cannot be dealt with using this decoupling method, given the available
interactions. To treat these cases we must introduce an additional
short-time assumption, which may or may not be more severe than the
usual BB assumption.

\subsection{Decoupling Method for System-Bath Interaction Without Symmetry}

Suppose again that $|J_{ij}|\gg |\Delta _{i}|\gg
|B_{i}^{x}|,|B_{i}^{y}|,|B_{i}^{z}|$, but that there is no symmetry in
the system-bath interaction, i.e., the case of arbitrary $H_{SB}=\sum_{i}\vec{\sigma }_{i}\cdot 
\vec{B}_{i}$.  Unlike the previous cases, where there was a
symmetry in $H_{SB}$, we now have to resort to a small time
approximation in order to expand the time evolution. Namely
\[
e^{i(A+B)}=\lim_{n\rightarrow \infty }\left( e^{iA/n}e^{iB/n}\right)
^{n}=e^{iA/n}e^{iB/n}+O(\frac{1}{n^2}).
\]
To see how this is useful, assume that we leave the system-bath
interaction on for a short time, so that: 
\begin{equation}
e^{-iH_{SB}t/n} = \prod_{i=1}^{N} e^{-i\sigma
_{i}^{x}B_{i}^{x}t/n} e^{ -i\sigma _{i}^{y}B_{i}^{y}t/n}
e^{-i\sigma _{i}^{z}B_{i}^{z}t/n} + O(\frac{1}{n^2}), 
\label{eq:exp}
\end{equation}
which can be partly decoupled using the Josephson interaction. First: 
\begin{eqnarray*}
C_{\sigma _{i}^{z}\sigma _{i+1}^{z}}^{\pi /2}\circ [ \prod_{\alpha
= x,y,z} e^{-i\sigma_{i}^{\alpha}B_{i}^{\alpha}t/n}]
= \\
e^{i\sigma
_{i}^{x}B_{i}^{x}t/n} e^{i\sigma _{i}^{y}B_{i}^{y}t/n}
e^{-i\sigma _{i}^{z}B_{i}^{z}t/n}.
\end{eqnarray*}
Therefore:

\begin{eqnarray*}
e^{-iH_{SB}t/n}\left[ C_{\sum_{i=1,3,...}^{N-1}\sigma _{i}^{z}\sigma
_{i+1}^{z}}^{\pi /2}\circ e^{-iH_{SB}t/n}\right] = \\
e^{-i\sum_{i}\sigma
_{i}^{z}B_{i}^{z}t/n}+O(n^{-2}). 
\end{eqnarray*}
The remaining term can be decoupled using the tunneling Hamiltonian: 
\[
e^{-i\sum_{i}\sigma _{i}^{z}B_{i}^{z}t/n}\left[ C_{\sum_{i}^{N}\sigma
_{i}^{x}}^{\pi /2}\circ e^{-i\sum_{i}\sigma _{i}^{z}B_{i}^{z}t/n}\right]
=I+O(\frac{1}{n^2}). 
\]
This procedure requires pulses that are short not only on the time
scale of the bath inverse frequency cutoff (for the BB procedure), but also on the time scale
of the system-bath interaction [in order to justify the expansion of
the exponential in Eq.~(\ref{eq:exp})]. We see once more that some decoherence
reduction can be performed without using more resources than is required for
computation.

\section{Conclusions}

By encoding two physical qubits into one logical qubit, or three
physical qubits into two logical qubits, we have shown how
some of the building blocks that have so far been considered
indispensable in superconducting phase-qubit quantum computers, can
in fact be eliminated. Moreover, by grouping the code's ancilla qubits into a single
``bus qubit'', we were able to further simplify the engineering
constraints on the fabrication of this important class of solid-state
qubits. An important additional advantage of the encoding is that it
allows one to eliminate potential external sources
of noise and decoherence. We have shown how to use
decoupling pulse methods in order to further drastically
suppress the remaining sources of errors
on these encoded qubits, without using extra space
resources. We believe that the approach presented here will prove to be
useful in reducing design constraints as well as decoherence and noise sources in
superconducting phase-qubit quantum computers.

\begin{acknowledgments}
We thank Dr. M.H.S. Amin, Dr. G. Rose, and especially Dr. A. Zagoskin for valuable
discussions. D.A.L. and L.W. acknowledge financial support from D-Wave
Systems, Inc. A.B. was supported by D-Wave Systems, Inc., NSERC, IMSI and FCAR.
\end{acknowledgments}

\end{document}